# THz-TDS time-trace analysis for the extraction of material and metamaterial parameters

Romain Peretti, Sergey Mitryukovskiy, Kevin Froberger, Aniss Mebarki, Sophie Eliet, Mathias Vanwolleghem, and Jean-François Lampin

*Abstract*—We report on a method to fit time-trace data from a terahertz time-domain-spectroscopy system enabling the extraction of physical parameters from a material or metamaterial. To accomplish this we developed a Python-based, open-source software, called Fit@TDS that functions on a personal computer. This software includes commonly used methods where the refractive index is extracted from frequency-domain data. This method has limitations when the signal is too noisy or when an absorption peak saturates the spectrum. Thus, the software also includes a new method where the refractive indices are directly fitted from the time-trace. The idea is to model a material or a metamaterial through parametric physical models (Drude Lorentz model and time-domain coupled mode theory) and implement this in the propagation model to simulate the time-trace. Then an optimization algorithm is used to retrieve the parameters of the model corresponding to the studied material/metamaterial. In this paper, we explain the method and test it on fictitious samples to probe its feasibility and reliability. Finally, we used Fit@TDS on real samples of high resistivity silicon, lactose and gold metasurface on quartz to show the capacity of the method.

## I. Introduction

The method of terahertz time-domain spectroscopy (THz-TDS), enabled by the progress in short-pulse lasers, begun rapid development about 30 years ago [1, 2, 3, 4]. This well-established technique is now mature and has been commercialized by several companies. Today, THz-TDS is the main tool for broadband terahertz (THz) spectroscopy, offering more than one-decade of bandwidth with a standard resolution up to ~ 1 GHz. THz-TDS has shown the capability to study different materials such as semiconductors [5], ferroelectrics, superconductors, liquids [6], gases [7, 8], biomolecules [9], and molecular crystal such as carbohydrates [10]. More recently, the measurements of band-pass filters [11] and metasurfaces embedded in microfluidic circuitry [12] were presented, along with more fundamental studies of, for instance, ultra-strong coupling [13, 14].

As its name suggests, this spectroscopic technique involves measurements made in the time domain, and does not utilize dispersive elements as in commonly used frequency-domain methods. Furthermore, in contrast to Fourier-transform infrared spectrometers where the measured function is the autocorrelation of the time-domain data through interferometry, THz-TDS is based on a direct measurement of the electric field in the THz frequency range.

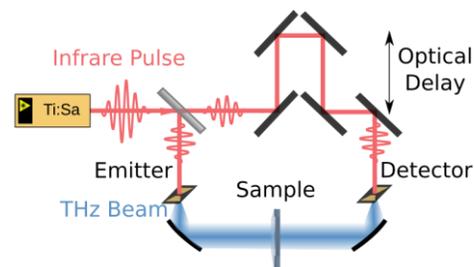

FIG. 1. Schematic of a typical THz-TDS experiment. It shows the femtosecond Ti:Sa laser exciting a photoconductive antenna that produces a THz pulse. This pulse travels through the sample and is then detected by another photoconductive antenna when illuminated with the femtosecond laser pulse after a controlled delay.

The working principle of a common THz-TDS setup is depicted in FIG. 1. A THz pulse is emitted by a THz antenna or a nonlinear crystal by means of the optical rectification effect of a near-infrared pulse produced by a femtosecond laser. Next, a lens or a parabolic mirror is used to collimate the pulse and to direct it toward the sample under study. The transmitted (or reflected) pulse is then collected by an optical system and aligned onto a detector. The detector measures the electric field versus time by means of photoconductive or electro-optical sampling with a typical time sampling between 10 and 50 fs. The ability to measure directly the electric field of the THz pulse rather than the averaged energy gives access to both the phase and the amplitude of the waveform, and thus provides information on the absorption coefficient and the refractive index of the sample, or on the

"This work was partially supported by: i) the international chair of excellence "ThOTroV" from region "Hauts-de-France" ii) the welcome talent grant NeFiStoV from European metropole of Lille and iii) the French government through the National Research Agency (ANR) under program PIA EQUIPEX ExCELSiOR ANR 11-EQPX-0015."

All Authors are with the Institut d'Electronique de Microélectronique et de Nanotechnologie (IEMN), CNRS, Univ. Lille, 59652 Villeneuve d'Ascq, France (corresponding author: romain.peretti@univ-lille.fr).

dispersion [15] in the case of a photonic element like a waveguide. This makes the THz-TDS method a powerful tool for characterization of materials and photonic devices. For material analysis, the usual way to retrieve material parameters is to perform a Fourier transform of the recorded pulse time-traces with and without a sample. The ratio between these two spectra is called the complex transmission coefficient and can be written as [16, 17]:

$$\tilde{T}(\omega) = \frac{\tilde{E}_s(\omega)}{\tilde{E}_{ref}(\omega)},$$

$$\tilde{T}(\omega) = \tilde{s}(\omega) \times exp\left(-j\frac{\omega d}{c}(\tilde{n}(\omega) - 1)\right) \times \widetilde{FP}(\omega) . \#\#(\text{Erreur ! Signet non défini.})$$

Here, $\tilde{E}_s(\omega)$ and $\tilde{E}_{ref}(\omega)$ are the Fourier transforms of time-domain signals $E_s(t)$ and $E_{ref}(t)$, respectively[1], $\tilde{n}$ is the complex refractive index where the real part corresponds to a delay and the imaginary part to an absorption in the material, $d$ is the thickness of the sample that must be measured, and $\omega$ is the angular frequency. The term $\tilde{s}(\omega)$ is the product of the Fresnel coefficients at normal incidence for the two air/material interfaces:

$$\tilde{s}(\omega) = \tilde{t}_{1\to2}(\omega) \times \tilde{t}_{2\to1}(\omega) = \frac{4\tilde{n}(\omega)}{(\tilde{n}(\omega) + 1)^2} . \#\#(\text{Erreur ! Signet non défini.})$$

Finally, $\widetilde{FP}(\omega)$ is a term taking into account the Fabry-Pérot multiple reflections in the sample [18]:

$$\widetilde{FP}(\omega) = \frac{1}{1 - (\tilde{r}_{2\to1}(\omega))^2 exp\left(-2j\frac{\omega d}{c}\tilde{n}(\omega)\right)} . \#\#(\text{Erreur ! Signet non défini.})$$

It has to be noted that using $\widetilde{FP}(\omega)$ as in eq. (3), may be a problem if not all the echoes above the noise floor of the TDS experiments are recorded (i.e. time-trace shorter than full signal). In fact, since the FFT algorithm performs a periodization of the signal in the time domain, in such a case the model will fold the echoes back to the beginning of the time-trace. To avoid such effect one can replace the expression in eq. (3) simply by the sum of the first terms of the FP series as shown in [19].

Equation (1) sets the so-called "forward problem": knowing $\tilde{E}_{ref}(\omega)$ and $\tilde{n}(\omega)$ one can obtain $\tilde{E}_s(\omega)$. Since the experiment gives $\tilde{E}_{ref}(\omega)$ and $\tilde{E}_s(\omega)$, the actual interest is the "inverse problem", that is, with knowledge of $\tilde{E}_{ref}(\omega)$ and $\tilde{E}_s(\omega)$, one can determine $\tilde{n}(\omega) = \tilde{\eta}(\omega) + \tilde{\kappa}(\omega)$. To the best of our knowledge, this problem can be solved analytically only by ignoring the Fabry-Pérot term (by, for example, using a temporal filter) and only for a sample without absorption. Since these assumptions imply that there is no phase term in the transmission coefficients in (2), the method consists of the extraction of the unwrapped phase from the THz-TDS data in the frequency domain, and then dividing by the frequency to obtain the refractive index. Another implication of these assumptions is that there are no losses during the pulse propagation—hence one can retrieve the transmission as a function of frequency and solve the second-order equation from (2) to retrieve the refractive index. As a consequence of the aforementioned assumptions, this method is limited only to optical-thick (*nd* > 1.5 mm), non-absorbing samples. Nevertheless, one can iterate this process by determining the real part of the refractive index from the unwrapped phase, and then compensate the difference of the losses by adding an imaginary term to the refractive index. Then, one has to compensate the phase term in the transmission due to this imaginary part—thus returning to the beginning of the loop. This iterative method is a good starting point; however, it does not guarantee the convergence or any reliability of the obtained results. Since this technique is intrinsically non-causal (the Kramers-Kroenig relation is not fulfilled), there is room for improvement by solving this inverse problem with a numerical approach. To do so, one should define an error function which must be minimized. This error function could be defined as [4]:

$$\xi(\tilde{n}(\omega))(\omega) = (\delta\rho(\omega))^2 + \psi(\delta\varphi(\omega))^2, \#(\text{Erreur ! Signet non défini.})$$

where $\delta\rho$ is the modulus error and $\delta\varphi$ is the phase error between the modelled transmission coefficient $\tilde{T}$ and the measured one $\tilde{T}_{meas}$. $\Psi$ is a weighting coefficient enabling the addition of the phase and modulus errors, its value is usually set to 1 *amplitude unit/rad*. The error function is defined for each frequency. Since the refractive index is related to the phase, especial care must be taken when calculating $\delta\varphi$. The measured phase is calculated as the unwrapped phase of $\tilde{T}_{meas}$ and the modelled phase is calculated as the following:

$$\arg(\tilde{T}(\omega)) = \arg\left(\frac{4\tilde{n}(\omega)}{(\tilde{n}(\omega) + 1)^2}\right) - \frac{\omega d}{c}(\tilde{n}(\omega) - 1) + \arg(\widetilde{FP}(\omega)). \#(\text{Erreur ! Signet}$$

Once the error function is defined, a minimization algorithm must be implemented. For example, one can use the simplex method (gradient free method) [20] or a quasi-Newton algorithm [21, 22]. The parameter search has to be done for every single frequency and gives as a result the real and the imaginary parts of the refractive index, $\tilde{\eta}(\omega)$ and $\tilde{\kappa}(\omega)$, respectively.

This method is fast and works regardless of whether the Fabry-Pérot effect is taken into account. However, the result does not respect causality (which takes the form of the Kramers-Kroenig relations in this problem). This creates a significant issue during the unwrapping step, which strongly depends on the dynamic range. It has been shown that it is possible to partially solve this problem by including a correction to the unwrapped measured phase using partial Kramers-Kroenig relations [23]. Moreover, in the error function (4), both the modulus and the phase errors have the same weight. This is arbitrary and any choice of weighting other than 1 *amplitude unit/rad* could improve or diminish the efficiency and accuracy of the algorithm.

Both the iterative and the optimization techniques show good results, but are limited—one still needs a precise measurement of the thickness or the implementation of an additional optimization step [24]. In addition, a low dynamic range of the measured data, or a strong absorption in the sample, leads to difficulties in obtaining the refractive index following the Kramers-Kroenig relations. This is due to the fact that the phase is lost in the frequency range where the signal value is below the noise level [23], which implies an additional step while performing the phase unwrapping. This phase includes additional assumptions in the number and the shape of the absorption peaks. Furthermore, the arbitrary weighting between the phase and the amplitude, as well as an arbitrary limitation of the bandwidth to avoid the range of low dynamic range, limits the robustness and thus the expansion of the optimization method.

---

[1] We took as a convention that all temporal functions are denoted with a letter and their Fourier transform with the same letter and a tilde.

Nevertheless, with the refractive indices retrieved by these techniques, it is possible to extract further information about the material itself. For this purpose, the refractive indices are fit using models such as the Drude-Lorentz model [18, 25]. One can then gain insight into the physical properties of the material under study, such as electronic or vibrational resonances. Knowledge of these parameters is of prime importance for material identification. This is, for instance, one of the most promising THz applications for drug component quality control or anti-counterfeiting measures. Most approaches focus on the intensity and the resonance frequencies while some take into account the linewidth [25]. Such parameter retrievals were achieved, for example, with the help of an optimization routine [26, 27], such as genetic algorithms [28], used in the frequency domain. Generally, the principle steps are the following: *(i)* performing the experiments with and without samples; *(ii)* computing the Fourier transform; *(iii)* measuring the thickness of the sample; *(iv)* extracting the real and/or imaginary part of the refractive index, and *(v)* fitting the refractive index to obtain the material parameters. This process has shown promising results, but has several drawbacks as previously described.

In this work, we present a robust and generic optimization approach [29, 30]. The Fit@TDS software is based on a direct comparison of the initial time-domain data of the measured THz pulse with a mode similarly to the work presented to measure thickness of paint layer by *Van Mechelen et al.* [19]. We introduce this software and show that it enables one to model both simple materials, such as silicon; as well as more complex ones, such as carbohydrates and even metasurfaces.

The remainder of this article is organized as follows. The basics of the implemented optimization is explained in section II. The two different models used are briefly described in section III. Sections IV and V present an analysis of the method's performance on fictitious and real samples, respectively.

## II. Optimization problem

The optimization problem is a problem of finding the best solution from all the feasible solutions. In the studied case, it starts with two items:

1. A set of data containing the time-traces with and without a sample (2 traces);

2. A model depending on the set of parameters $\{p_i\}$ depicting how the sample transforms the reference pulse into the modeled one, $E_{model}\{p_i\}(t)$.

Concretely, an example of a model for a doped semiconductor sample measured in transmission would transform $E_{ref}(t)$ to $E_{model}\{p_i\}(t)$ simply by convoluting $\tilde{E}_{ref}(\omega)$ with $T\{p_i\}(t)$, calculated by introducing the Drude-Lorentz model equation in the refractive index in (1) (see Sec. III for details). Then, the objective function to minimize is set as the L² norm (square root of the sum of the square of the differences) of the difference between the modelled pulse and the measured (sample) one:

$$Obj\{p_i\} = \sum_{t=0}^{t=t_{max}} \left(E_{model}\{p_i\}(t) - E_s(t)\right)^2 dt \ . \ \#\#(\text{Erreur ! Signet non d}$$

This function will vary upon the value of parameters of the chosen model and the goal of the optimization is to determine the set of parameters that minimizes the objective function. An important remark here is that the function we are minimizing is proportional to electromagnetic energy. The fact that the L² residual error is an intuitive physical quantity will help the user in interpreting the results, and to understand any discrepancies in either the experiment or the model. This will, for instance, facilitate the understanding of any divergence or convergence of the fit algorithm to some local minimum, or to (in)validate the choice of the model during the optimization. The other practical advantage of this formulation is given by Parseval's theorem, which states that the norm of a function is the same as the norm of its Fourier transform, meaning:

$$Obj\{p_i\} = \sum_{\omega=-\omega_{min}}^{\omega=-\omega_{max}} \left|\tilde{E}_{model}\{p_i\}(\omega) - \tilde{E}_s(\omega)\right|^2 d\omega \ . \ \#(\text{Erreur ! Signet no})$$

This is extremely convenient, allowing the calculation of the objective function in both time and frequency domains without performing a Fourier transform at each iteration. Specifically, since, to our knowledge, all the refractive index models are defined in the frequency domain, this formula allows one to perform the time-domain optimization while computing the objective function in the frequency domain.

## III. Models

To perform the optimization, we coded the methods as given in repository from Ref. [31] in a mainstream language (Python) that allows the use of different optimization libraries. Since our goal is to offer a broad tool to the community, that is usable regardless of the type of sample under study or the subsequent problem to solve, we implemented a function from a research library [32] that includes many different optimization algorithms—giving users the opportunity to choose the ideal one for their samples. For the choice of the algorithm itself, we chose the versatile augmented Lagrangian particle swarm optimizer

(ALPSO) [33]. This algorithm is an improvement on the basic "swarm particle optimization" approach used in constrained engineering design and the optimization problem.

We implemented two different models: one for solid materials and one for metamaterials. In both cases, we assume that the sample is made only of one layer—meaning two interfaces added to a propagating medium.

*A. Multiple Drude-Lorentz Model*

Bulk solid materials play a role in the pulse propagation simply due to their refractive index in equation (1) and in the Fabry-Pérot term therein. We implemented the Drude-Lorentz model (the most commonly used model for solid samples), which defines the dielectric permittivity of a sample as a set of electronic resonators (matrix vibrations, oscillating charges, etc.) and leads to the following permittivity function:

$$\tilde{\varepsilon}(\omega) = \tilde{n}^2(\omega) = \varepsilon_\infty + \frac{\omega_p^2}{\omega^2 + j\omega\gamma_p} + \sum_{k=1}^{k_{max}} \frac{\Delta\varepsilon_k \omega_{0,k}^2}{\omega_{0,k}^2 - \omega^2 + j\omega\gamma_k}, \text{\#(Erreur ! Signet non défini.)}$$

where $\varepsilon_\infty$ is the dielectric permittivity at high frequency compared to the range of interest, $\omega_p$ is the plasma frequency, $\gamma_p$ is the damping rate, $k_{max}$ is the number of considered oscillators, $\omega_{0k}$, $\gamma_k$ and $\Delta\varepsilon_k$ are the resonant frequency, the damping rate and the strength (expressed in permittivity units) of the $k^{th}$ oscillator, respectively. This formula will be useful to model the phonon line in a semiconductor or in a molecular crystal in the THz range.

To summarize, in the case of a single uniform layer, the propagation will be modeled through the Fresnel coefficients and the multiple Drude-Lorentz oscillator model using $3 \times [k + 1]$ parameters.

*B. Metasurfaces*

For a metamaterial, the model is different because one has to take into account (i) the refractive index of the material used to build the metamaterial, and (ii) the interaction of the THz pulse with the metamaterial structure itself. As often in any physical model, one can choose a macroscopic approach or a microscopic one. Since a metamaterial is a macroscopic concept the first approach is the most used: one model the metamaterial in terms of effective refractive index (permittivity and permeability …). Such an approach is especially useful for metamaterial based applications but has the drawback of being less intuitive when it comes to optimization of the design or the fabrication. A time-domain coupled-mode theory (TDCMT) model [34] of the resonator constituent of the metamaterial has been adopted so as to provide more insight in the physics of the device. This approach has been used, for instance, to model an integrated photonic resonator [34] or, in the free-space case (which is closer to the scope of this work), a photonic crystal in a solar cell [35]. Thus, it is fully appropriate for a metasurface built out of resonators, such as split ring resonators (SRR). We derived a similar equation as found in Ref. [34] with the addition of the reflection and transmission at the air/sample interfaces, and the result gives:

$$\tilde{r}_{\text{meta}}(\omega) = \frac{S_{-1}(\omega)}{S_{+1}(\omega)}$$
$$= \tilde{r}_{2\to 1}(\omega) - \frac{exp(j\delta\theta)}{\sqrt{\tau_{e2}\tau_{e1}}\left(j(\omega - \omega_0) + \frac{1}{\tau_0} + \frac{1}{\tau_e}\right)},$$

$$\tilde{t}_{\text{meta}}(\omega) = \frac{S_{-2}(\omega)}{S_{+1}(\omega)}$$
$$= \tilde{t}_{1\to 2}(\omega) - \frac{1}{\tau_{e1}\left(j(\omega - \omega_0) + \frac{1}{\tau_0} + \frac{1}{\tau_e}\right)}, \text{\#(Erreur ! Signet non défini.)}$$

where $\tilde{r}_{2\to 1}(\omega)$ and $\tilde{t}_{1\to 2}(\omega)$ are the reflection and transmission coefficients (shown in FIG. 2) following the Fresnel law, $\omega_0$ is the resonant frequency of the mode, $\tau_0$ is the characteristic time for absorption losses, and $\tau_e$, $\tau_{e1}$, and $\tau_{e2}$ are the characteristic times for external losses in total, toward direction 1 (incoming) and toward direction 2 (outgoing), respectively.

Note that an additional hypothesis arises here: we assume that the metal of the metasurface does not influence the transmission or reflection outside of the resonance spectral ranges—meaning that the filling factor (area ratio) of the metal must be very low.

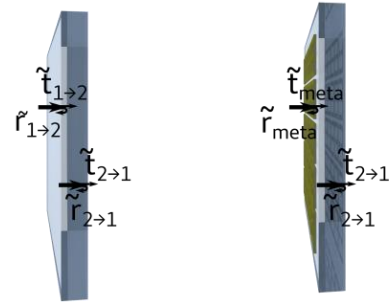

FIG. 2 Schematic showing the Fresnel coefficients for the simple layer (left), and the metasurface (right) as in equation (9).

Equation (9) will play a role in changing the transmission term:

$$\tilde{s}(\omega) = \tilde{t}_{\text{meta}}(\omega) \times \tilde{t}_{2\to 1}(\omega) = \tilde{t}_{\text{meta}}(\omega) \frac{2\tilde{n}(\omega)}{(\tilde{n}(\omega) + 1)}, \text{\#\#(Erreur ! Signe}$$

and in the Fabry-Pérot term can then be written as:

$$\widetilde{FP}(\omega) = \frac{1}{1 - \tilde{r}_{\text{meta}}(\omega)\tilde{r}_{2\to 1}(\omega)exp\left(-2j\frac{\omega d}{c}\tilde{n}(\omega)\right)}. \text{\#\#(Erreur ! Signe}$$

To summarize, in the case of a metasurface, the substrate will be modelled using the same Drude-Lorentz parameters with five additional parameters to take into account the resonant nature of the transmission and the reflection at the metasurface interfaces (frequency and, internal and external losses).

It has to be noticed that this model is well suited for infinitely thin metasurface as those based on metallic structures. In the case of relatively thick dielectric metasurfaces [36, 37] one would have to modify the model to introduce the two interfaces before and after the metamaterial layer.

*C. Implementation*

The proposed models can involve more than thirty parameters for complex samples, thus an exhaustive error calculation to reach the global maximum for each set of parameters is too demanding. As stated above, the strategy we use is to implement an optimization algorithm to solve this nonlinear problem. A tremendous number of algorithms have been developed—giving birth to several fields of research. We implemented a library offering several optimization algorithms (the Python-based optimization package called PyOpt [32]), which is designed to formulate and solve nonlinear constrained optimization problems. The main advantage of this package is that it includes 20 different optimization algorithms, allowing users of Fit@TDS to change the algorithm for one that is more efficient for his or her specific problem. In addition, PyOpt allows parallelization, which is extremely useful for diminishing computation time. Since the problem is to find a global maximum rather than a local one, we implement the optimized particle swarm routine ALPSO amongst the proposed algorithms. The particle swarm algorithm is a versatile meta-heuristic method that makes no or few assumptions about the problem being optimized, and can search very large spaces of candidate solutions [38, 39]. The drawback of this choice is that it is not optimized in terms of computation time for our specific problem(s). Of course, this could be improved in the future with better-suited algorithms.

## IV. VALIDATION AND PERFORMANCE ASSESSMENT WITH SIMULATED FACTITIOUS SAMPLES

To validate and assess the performance of the proposed methods we first used a simulated sample. To do so, we recorded the reference spectrum with the TeraSmart THz-TDS spectrometer by Menlo Systems GmbH (1000 accumulations) [40] and then numerically simulated the response of the system with a sample using the equations described above. The data without a sample were windowed at the end of the time-trace on a segment length corresponding to the time delay introduced by the sample to remove folding effects due to the periodicity of the FFT. For the data with the simulated sample, we convolved the measured pulse by the transfer function of equation (1) with the Drude-Lorentz model to define the permittivity. Then, we added Gaussian white noise to the time-trace with a magnitude equivalent to the level of the high frequency noise one can observe in FIG. 3 ($5\times10^{-5}$ of the amplitude unit used, and approximately -90 dB compared to maximum power spectral density). We used these two sets of time-trace data as the input to Fit@TDS.

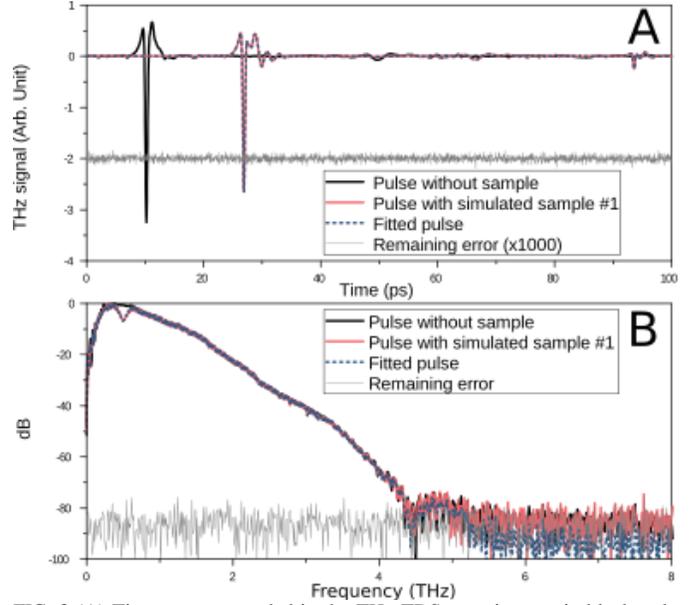

FIG. 3 (A) Time-traces recorded in the THz-TDS experiments: in black – the reference, in red – simulated time-trace from one oscillator sample, in green - result of the fit, in gray - the difference between the fit and the simulated data (multiplied by 1000); (B) the corresponding spectra.

*A. First validation: one oscillator Drude-Lorentz model*

To perform a first validation we tested Fit@TDS with a simulated 5-mm-thick sample with a dielectric constant build with the one-oscillator Drude-Lorentz model with the following parameters: $\varepsilon_\infty = 4$; $\Delta\varepsilon = 0.01; \omega_0 = 0.5 THz$; $\gamma = 0.1 THz$. The time-trace of the sample is shown in FIG. 3, and the resulting permittivity is plotted in FIG. 4.

The spectral data clearly features a dip at the frequency of the oscillator. Then, we implemented our software Fit@TDS, with the ALPSO algorithm using the swarm-size of 1000, 6 inner iterations and 20 outer iterations. The bounds for the thickness were ±1% around 5 mm. The bounds for the four other parameters were − 50% and +100% of the parameter value. Although we optimized neither the choice of the algorithm, nor the parameters (swarm size, inner and outer iterations, etc.), the software required about 1 minute on a common personal computer (Intel® Core™ i7-5600U CPU @ 2.60 GHz) to retrieve the parameters from a 300-point data set. In fact, the computation time strongly depends on the size of the parameters space, and thus on the bounds we specify. In the difference between the fit and the simulated data shown in FIG. 3, one can see an extremely small discrepancy between the targeted time-trace and the fitted one. This discrepancy comes from the noise added to the simulated sample time-trace, meaning that the algorithm converged and that the $L^2$ residual error is simply the $L^2$ of the added noise. The results for the Drude-Lorentz parameters are very close to the targeted ones, the relative error is $E(\varepsilon_\infty) = 10^{-7}$; $E(\Delta\varepsilon) = 6\times10^{-5}$; $E(\omega_0) = 8\times10^{-6}$; $E(\gamma) = 8\times10^{-5}$ and $1\times10^{-7}$ for the thickness discrepancy.

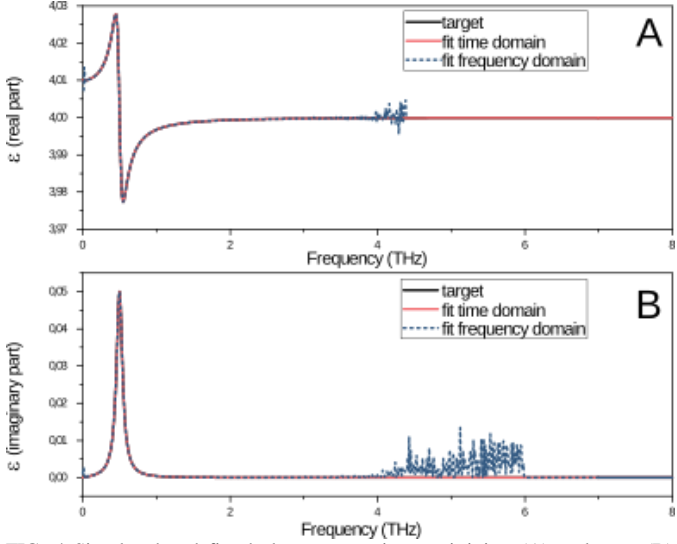

FIG. 4 Simulated and fitted electromagnetic permittivity: (A) real part, (B) imaginary part.

The error of the thickness should be compared to the delay line uncertainty:

$$\delta d = \frac{c\delta t}{\eta} \sim 9 \times 10^{-7}\, m. \#(\text{Erreur ! Signet non défini.})$$

The errors on $\varepsilon$ and $\Delta\varepsilon$ are to be compared to:

$$\delta\varepsilon = 2\left(\frac{c\delta t}{d}\right)\sqrt{\varepsilon} \sim 7 \times 10^{-4}, \#\#(\text{Erreur ! Signet non défini.})$$

Which correspond to the sampling time of our THz-TDS setup. The errors on $\omega_0$ and $\gamma$ are to be compared to the frequency sampling:

$$\delta\omega = \frac{2\pi}{\Delta T} \sim 6 \times 10^{10}\, rad/s. \#(\text{Erreur ! Signet non défini.})$$

All results show that for a Lorentz-modelled time-trace that is not limited by noise (see Eq. 15), the proposed methods and the software are validated and reach the target precision.

### B. Dependency of the error on the dynamic range

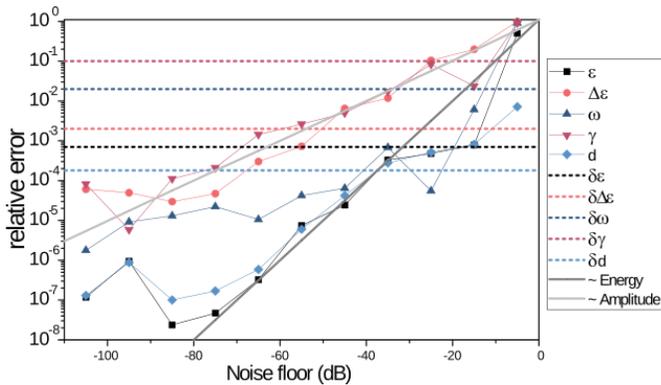

FIG. 5. Relative error on the parameter retrieval versus applied noise level of the measurements Eqs. 12, 13 and 14. The dotted lines depicts the usual comparison points in term of precision.

To better evaluate the actual performance of the method on real data, we tested its robustness by increasing the amount of noise added to both time-traces with and without a sample. Specifically, the amplitude of the Gaussian white noise was increased from $5\times10^{-5}$ up to 5 amplitude units, corresponding to a dynamic range of 105 dB to 5 dB. Then, the algorithm was used to retrieve the parameters using the same bounds as in the previous case. The relative error of the parameters as a function of the applied noise level is shown in FIG. 5. The $L^2$ residual error was not shown, since it always corresponds to the added noise (written here as the noise floor).

Firstly, one can see that, globally the errors increase with the applied noise, as expected. Secondly, the relative error follows a trend between proportional to the amplitude and proportional to the energy, as indicated by the solid gray lines in FIG. 5. Thirdly, the relative uncertainty is larger for $\gamma$ and $\Delta\varepsilon$ simply because those parameters are intrinsically smaller. Finally, one can compare the retrieval precision with the value mentioned above and plotted in dotted lines on the same figure. Clearly, when the noise is low, we exceed the targeted precision for each parameter (i.e. the software yields better-than-expected results). Furthermore, we reach a precision below 1% for a noise floor of -40 dB. This level of noise is in fact the "single shot" noise in our experiments (meaning one scan by the THz-TDS delay line taking ~ 50 ms of acquisition time). This result implies that the method is robust enough to follow the parameters of the one-oscillator Drude-Lorentz model in real time with an data with a noise level corresponding to experiments at 20 Hz repetition (video frame time). This will enable one, for instance, to follow the temporal evolution of a parameter extracted from the model (typically the width of an oscillator may depend on temperature) at this time frame.

### C. Resolution test

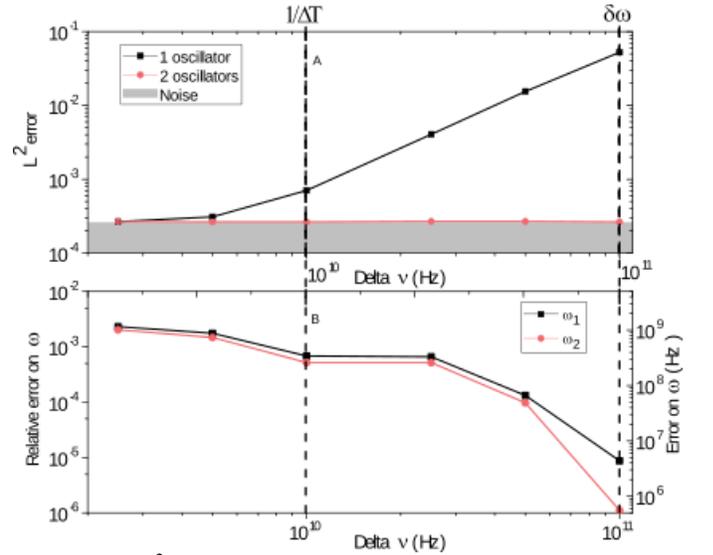

FIG. 6. (A) $L^2$ residual error of the fit using one or two oscillators (gray area depicts the limitation due to the added noise). (B) Relative error on the $\omega_1$ and $\omega_2$ parameters. Here, one can see that we are not limited by the width of the resonance but by the duration of the time-trace

The previous results show a very accurate retrieval of the frequency parameters even in the presence of strong noise. However, this does not mean that one would be able to discriminate a doublet: two different peaks separated by a frequency that is small compared to their resonant frequency $v_1$ and $v_2$ (resolution). The criterion that determines whether we are able to discriminate between two neighbouring peaks is simple: the $L^2$ residual error of the fit given by a two-oscillator model must be significantly smaller than that of a one-oscillator model. To test the performance of our methods regarding this criterion, we simulated a fictitious sample with the same parameters as in section A, on which we added an oscillator at slightly higher frequency. Then, we fit the data coming from this fictitious sample using first one oscillator and then two oscillators. The results as a function of the separation between the two oscillators, $\delta v$, are shown in FIG. 6.

One can see that the resolution of our fit is clearly below the width, $\delta \omega$, of the peaks themselves. However, we are limited by the temporal window of the experiments (100 ps). With the time-domain fit, a doublet will give a beat note, meaning a low frequency envelop on a high frequency carrier wave. Thus, if the dynamic range is high enough, the only limitation will be the time window of the experiment, which should be long enough to detect the variations of the envelope. Since the peaks have a finite size, the envelope will also be damped. Consequently, it will be lower than the noise. From these considerations, one can derive the following equation giving the optimal time window:

$$\Delta T_{opt} = \frac{2\pi}{\gamma} \ln\left(\frac{\tilde{A}\gamma}{2\pi \tilde{N} \delta f}\right)$$
$$= \frac{2\pi}{\gamma}\left[\ln\left(\frac{\gamma}{2\pi \delta f}\right) + \ln\left(\frac{\tilde{A}}{\tilde{N}}\right)\right](\delta\varepsilon, \gamma, N)$$
$$= \frac{2\pi}{\gamma} \ln\left(\frac{\tilde{A}\gamma}{2\pi \tilde{N} \delta f}\right) = \frac{2\pi}{\gamma}\left[\ln\left(\frac{\gamma}{2\pi \delta f}\right) + \ln\left(\frac{\tilde{A}}{\tilde{N}}\right)\right], \#(\text{Erreur ! Signet non défini.})$$

where $\tilde{N}$ is the noise amplitude value in the frequency domain, $\delta f$ is the full frequency range, and $\tilde{A}$ is the depth of the considered peak in the frequency domain. It is important to note that $\tilde{A}$ and $\tilde{N}$ are power spectral densities and thus are expressed in "squared-amplitude per Hertz" units. Finally, it is important to note that the term $\delta f$ is in fact the inverse of the sampling time. To summarize, as expected, the resolution of the proposed method is higher than the width of the peak and is still is limited by the frequency resolution of the discrete Fourier transform.

*D. Validation with multiple-oscillator Drude-Lorentz model*

|   | Targeted value | | | Relative Error (%) | | |
|---|---|---|---|---|---|---|
| # | $\Delta\varepsilon$ | $\omega/2\pi$ (THz) | $\delta\gamma/2\pi$ (THz) | $\Delta\varepsilon$ | $\omega/2\pi$ | $\delta\gamma/2\pi$ |
| 1 | $10^{-3}$ | 0.3 | 0.1 | 0.03 | 0.003 | 0.1 |
| 2 | $2.10^{-3}$ | 0.5 | 0.01 | $10^{-3}$ | $3.10^{-5}$ | 0.004 |
| 3 | $10^{-2}$ | 1 | 0.5 | 0.1 | 0.008 | 0.02 |
| 4 | $10^{-2}$ | 1.1 | 0.55 | 0.1 | 0.01 | 0.05 |
| 5 | 0.1 | 2.5 | 0.1 | 0.01 | 0.003 | 0.04 |
| 6 | $10^{-3}$ | 3.5 | 0.1 | 0.8 | 0.03 | 3 |

TABLE 1 : TARGETED VALUES FOR THE FIT AND CORRESPONDING RELATIVE ERRORS.

To validate the method with a more complex sample, we simulated one fictitious sample (like a carbohydrate) made of six oscillators with parameters given in Table 1. These parameters lead to a refractive index depicted in FIG. 7 and subsequently the spectrum shown in FIG 8.

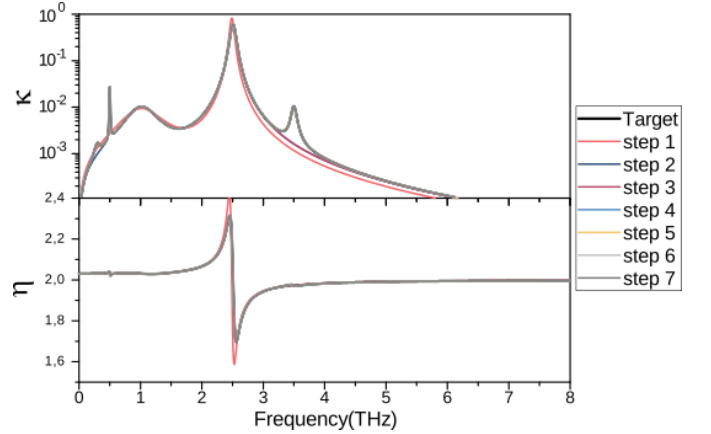

FIG. 7 Targeted refractive index versus frequency compared to the one calculated at each step of the fit (additional oscillators).

From the spectrum, one can see four dips. An additional absorption feature can be seen in the retrieved refractive index around 0.5 THz, corresponding to a fifth dip. If one intends to directly find the result, five dips requires 18 fit parameters, which corresponds to an enormous parameter space (e.g. taking only 10 values for each parameter would mean $10^{18}$ sets to test). Consequently, we began by fitting the two strongest oscillators (the one leading to the dip slightly above 1 THz and the one around 2.5 THz) and then adding the next oscillators one by one to strongly diminish the volume of possibilities. The new oscillators were added by picking the ones corresponding to the most significant residual error in the spectrum [as shown in Fig. 8 (B)]—leading to a decrease of the error as shown in Table 2.

| Step # | 1 | 2 | 3 | 4 | 5 | 6 | 7 |
|---|---|---|---|---|---|---|---|
| Number of oscillators | 2 | 3 | 4 | 5 | 6 | 6 | 6 |
| $L^2$ Residual error in % | 11 | 0.8 | 0.23 | 0.19 | 0.028 | 0.0269 | 0.0268 |

TABLE 2 : $L^2$ RESIDUAL ERROR DEPENDING ON THE NUMBER OF OSCILLATORS IN THE FIT.

Here, one can see that the residual error decreases step by step with additional oscillators. Moreover, in FIG. 8 (B) it is clear that the addition of a new oscillator decreases the residual error in the frequency range where the oscillator is added. After step #4, a residual error remains in the region around 1 THz. Thus, we added a sixth oscillator that strongly decreased the residual error. This is due to the fact that, in this region, two oscillators were present in the model (Table 1). Finally, because convergence was not fully obtained, we performed the sixth and seventh steps to refine the precision by fitting in a more constrained parameter space than used previously. This resulted in a precision and resolution similar to the ones obtain in FIG. 5 for each oscillator. Indeed, the higher the oscillator frequency, the lower the signal, and therefore the lower the precision. These results show that the method is not only valid for one or two oscillators but for a set of up to six oscillators, even if two of those oscillators are close together in frequency similarly to the example of FIG. 6.

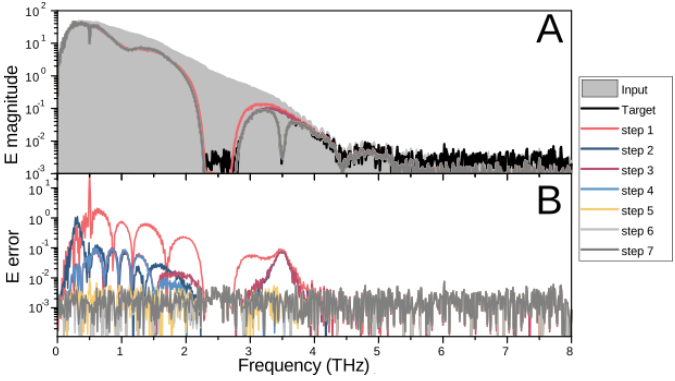

FIG. 8 (A) Targeted spectrum compared to the one fitted at each step, in gray - the spectrum of the input. (B) $L^2$ residual error spectrum at each step.

To summarize the first validations on fictitious samples, we proved that Fit@TDS enables the retrieval of the one-oscillator Drude-Lorentz model very accurately, even in a presence of a noise comparable to a single-shot video time-frame measurement in a commercial THz-TDS system. We showed that the proposed method is capable of identifying two different peaks of the two-oscillator Drude-Lorentz model with a resolution power that depends on the total time span of the recording, and not on the width of the peaks (as long as the measurements are not noise limited). We derived an expression (Eq. 15) to predict the optimum time span for THz-TDS measurements. Finally, we tested Fit@TDS on an example sample including six oscillators adopting an iterative use of the method. The results we obtained were at the same precision and resolution as those for a fewer number oscillators. The recursive approach we used for this fictitious sample allowed keeping a reasonable computational load while adding up to seven oscillators. The idea is to use the information remaining in the residual error to improve the model by adding oscillators in the vicinity of a clear bump in the residual error.

Nevertheless, it is important to test our method on real samples. For instance, with such samples, we are given the ultimate precision of the sample thickness, as well as the other parameters. In real cases, the thickness will be measured at some relative precision that may vary along the surface of the sample. In addition, the other parameters must be inferred from experiments and not from an artificial input.

## V. VALIDATION WITH REAL SAMPLES

After testing the method on fictitious samples, it was used on real samples. First, on two high-resistivity 5-mm-thick silicon wafers, then on a lactose pellet, and finally on a metasurface made of gold split-ring resonators on a quartz substrate.

### A. Silicon wafer sample

High-resistivity silicon is a typical reference material for THz-TDS. Several historical studies have been published since the emergence of the THz-TDS field [4, 41, 16, 42], hence this type of sample is a good starting point to test our method. A float-zone high-resistivity (> 10 kΩ·cm) silicon wafer with a thickness of 5 mm ± 1% was purchased from Sil'tronix Silicon Technologies. We measured the thickness with a digital thickness comparator from Mitutoyo to be 5016 ± 4 µm. We note that this value is likely overestimated, since any imperfections on the wafer or dust between the wafer and the marble plate will create additional thickness. We performed the THz-TDS measurement at a temperature of 23 ± 1°C, in a timing window of 560 ps with steps of 33 fs. Before plotting and treating the data, we bandpass filtered them with a box car frequency profile below 160 GHz to remove spurious parasitic modes. The time-traces and the corresponding spectra are presented in FIGS. 9 and 10, respectively.

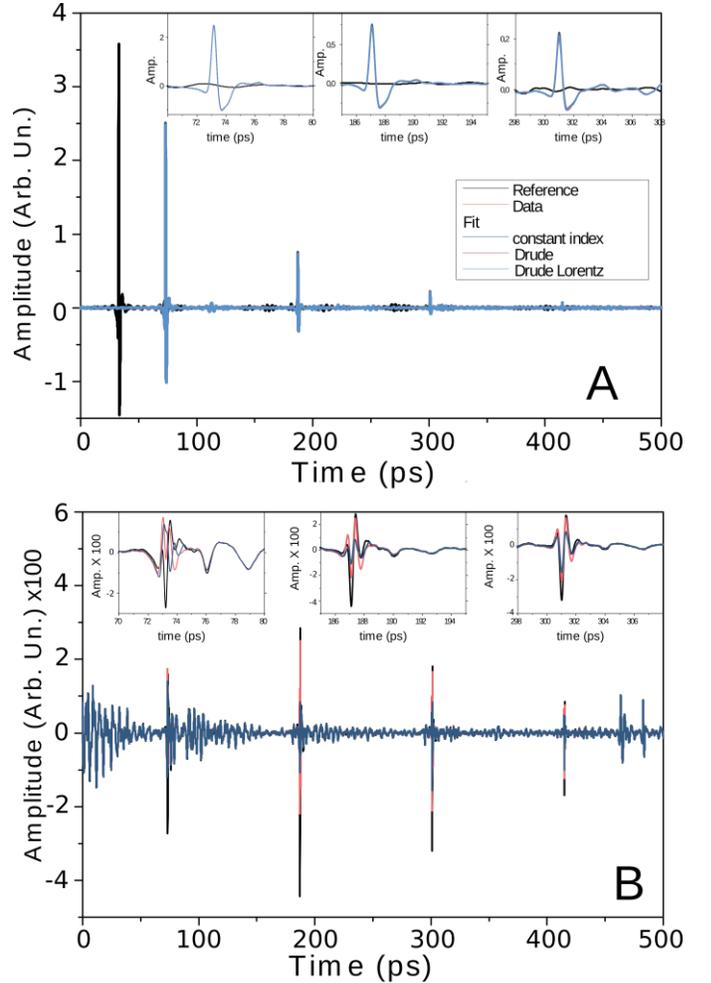

FIG. 9 (A) Time-trace of the Si wafer experiments (inset – zoom on the main peak and the two first echoes). (B) Residual error in the time domain.

Such samples are usually modeled using the Drude model [43]. Consequently, we fitted the data using first a non-dispersive permittivity (constant index), then a pure Drude model, and finally the Drude-Lorentz model with an oscillator to take into account the low energy branch of phonons [44, 45]. The complex refractive indices for all three models are plotted in FIG. 11 and the corresponding parameters in Table 3.

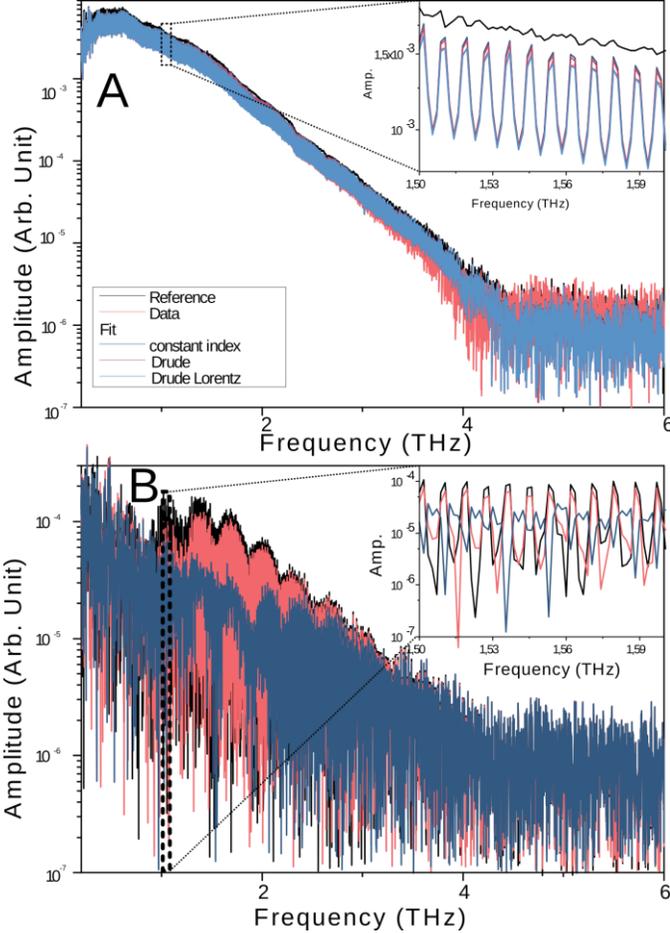

FIG. 10 (A) Spectra corresponding to time-trace of FIG. 9 (inset – zoom between 1.5 and 1.6 THz). (B) Residual error in the frequency domain.

Firstly, from FIGS. 9 and 10, we conclude that the fit worked fairly well for all three models, as is confirmed by the small value of residual error listed in Table 3. Secondly, one can see in the time domain that the temporal position of the residual error is clearly physical and thus can be interpreted. The first residual error from 0 ps is an artifact due to the box car filtering, and can be seen at lower magnitudes close to the end of the time-trace. Then the main source of residual error is perfectly time-correlated with the recorded pulses, and does not show any specific spectral feature—meaning that the models do not perfectly fit the experiments. This is the case for constant index and Drude model since one can see a real improvement by implementing the Drude-Lorentz model. In this case, the fact that the frequency domain residual error follows the frequency shape of the pulse fairly well, and that an oscillation at the Fabry-Pérot period is observed in the frequency domain, may indicate again that the model may not be fully sufficient. In fact, more sophisticated models have already been drawn for silicon in the THz domain, as in Ref. [46], for example, which is based on microscopic transport on which one can add a Lorentz oscillator to take into account low-energy phonons or absorption due to impurities.

To go a step further, the refractive indices retrieved by time-domain and frequency-domain methods were plotted in FIG. 11. We empathize that the thickness used for the fit in the frequency domain was the one extracted from the fit in the time domain (4.99908 mm), which is in good agreement with the comparator measurement. We also performed the fit with the measured thickness, but obtained a shift of -0.01 in the refractive index, and more importantly additional noise (which manifested as oscillations in the refractive index of amplitude ~$1.5 \times 10^{-3}$). These facts indicate that the measured thickness for three points on the wafer does not correspond to its average thickness. Here, it is clear that the time domain model is in total agreement with the frequency domain model—illustrating the coherence between the two methods. Moreover, our method is very precise and yields the effective thickness of the sample, which is extremely important for the study.

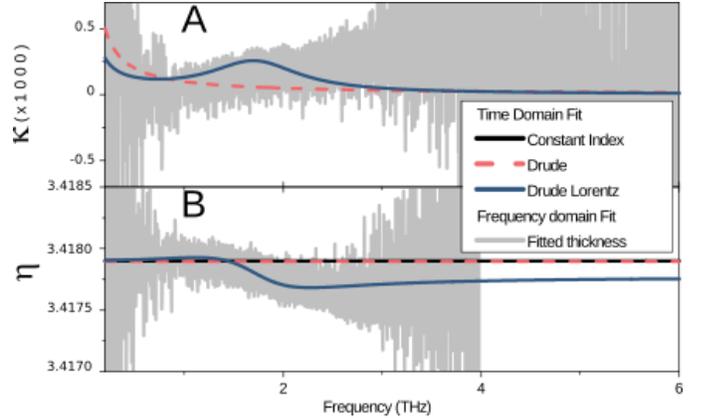

FIG. 11 Fitted refractive index of the two Si wafers modelled without oscillators, and with a Drude oscillator. The frequency domain fit was done using the thickness obtained from the time domain fit.

As for the fit results using the Drude-Lorentz model, they agree with the literature on the global value for refractive index [16]. Additionally, both the frequency and the width of the Lorentz oscillator tend to be very close to the one reported in Ref. [45]—showing the reliability of the method.

To conclude, we were able to predict the refractive index and absorption of a high-resistivity silicon wafer with a high precision. Considering the effect of jitter in the time-domain measurements in Eq. (13), and the fact that we did not measure the refractive index of nitrogen gas in the band that may induce systematic error, we trust the three first digits of this refractive index. Exceeding this level of accuracy would require additional stabilization (e.g. in temperature) and measurements that are beyond the scope of this paper. Nevertheless, we obtained values very close to the metrological one in the literature with a faster and simpler method, which is more than sufficient for the majority of applications.

| Model | No dispersion | Drude | Drude Lorentz |
|---|---|---|---|
| Residual Error % | 2.736 | 2.527 | 2.405 |
| T (mm) | 4.99908 | 4.99909 | 4.99907 |
| $\varepsilon$ | 11.68204 | 11.68201 | 11.68113 |
| $\omega_p/2\pi$ (THz) | | 4.19 | 18.82 |
| $\gamma_p/2\pi$ (THz) | | 25 700 | 964 617 |
| $\Delta\varepsilon$ | | | 0.00093 |
| $\omega_0/2\pi$ (THz) | | | 1.79 |
| $\gamma_0/2\pi$ (THz) | | | 1.10 |

TABLE 3 : SILICON REFRACTIVE INDEX PARAMETERS FOR THE THREE MODELS OBTAINED BY THE TIME DOMAIN FIT.

### B. Lactose pellet

If silicon is a perfect first example, it does not have many features to be fitted in the THz range. Thus, we used the methods on a pellet of lactose monohydrate (CAS 5989-81-1) powder with purity ≥ 99% total lactose basis (determined by gas chromatography) purchased at Sigma-Aldrich. Due to the fragility of the pellet, it was extremely difficult to measure the thickness. Nevertheless, we measured roughly 900 µm with an uncertainty of ± 20 µm. To fit the data, we followed the methodology used in section IV. We first analyzed the transmission data and found two strong peaks and numerous Fabry-Pérot oscillations. Hence, we fit the data using the two-oscillator model. The resulting $L^2$ residual error exhibited two additional oscillators, as well as some losses at high frequency. Consequently, we added three oscillators to take into account these features. The fit results are shown in FIG. 12 superimposed with frequency-domain fitted refractive indices. Table 4 lists the resulting fit metrics.

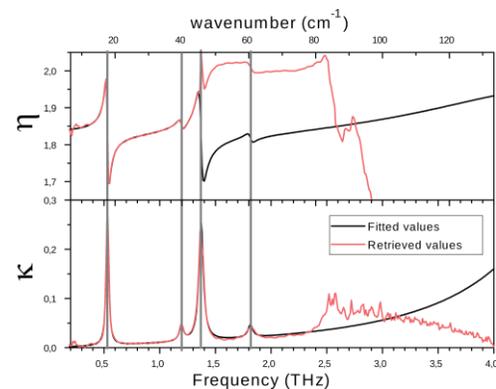

FIG. 12 Fitted value with the time-domain method in green and frequency domain fit in gray. The time-domain method avoids the problem linked to the loss of the phase in the frequency domain due to a too strong absorption.

Again, because the thickness predicted by the time-domain fit was more reliable than that obtained with comparator measurements, we used the former to retrieve the refractive indices in the frequency domain. Frequency domain and time domain indices are in good agreement at low frequency. However, from the highest peak around 1.37 THz up to the end of the spectral band, an important discrepancy was found due to a strong absorption peak. Evidently, the phase is lost in the frequency domain and thus the algorithm is not able to unwrap it properly [23]. This example clearly shows the robustness of the proposed method considering this issue. To go a step further, one can see that three absorption peaks are distinguishable with corresponding a feature in the real part of the refractive index. The peaks at 0.53 THz [47, 48], 1.17 THz [49, 50] and 1.37 THz [51, 52] are the characteristic absorption peaks of α-lactose monohydrate [10, 49, 53]. The peak at 1.81 THz is more difficult to measure (higher frequency and in the vicinity of a water line), and is therefore less prominent in the literature. However, it was reported in Ref. [52] and may be due to the presence of an anhydrous phase [54].

Additionally, the retrieved value for $\eta$ becomes unreliable after the strongest absorption peak at 1.37 THz. At this frequency, the signal decreases to the level of the noise, and therefore the phase becomes extremely noisy—leading to an error in the phase unwrapping. Finally, the shape of high-frequency losses does not match the shoulder of the hypothetical strong-infrared Lorentzian absorption. Thus, we tentatively attributed this last feature to losses due to the scattering in the pellet made of nano-crystallite powder.

| # | $\Delta\varepsilon$ | $\omega_0/2\pi$ THz / cm$^{-1}$ | $\gamma_0/2\pi$ GHz |
|---|---|---|---|
| 1 | 0.052351 | 0.5303 / 17.69 | 25.8 |
| 2 | 0.031530 | 1.3699 / 46.69 | 47.8 |
| 3 | 0.004434 | 1.1951 / 39.86 | 45.2 |
| 4 | 0.002738 | 1.8137 / 60.50 | 54.1 |
| *5* | *0.509754* | *5.0764 / 169.3* | *1618* |

TABLE 4 : RESULTS OF THE FIT OF THE LACTOSE PELLET WITH FIVE OSCILLATORS. THE FIFTH OSCILLATOR IS USED TO FIT SCATTERING LOSSES.

To summarize, we were able to retrieve the frequency, the width and the oscillator strength of four oscillators in the range of 0.2 to 2.5 THz. Further improvements could be made by including scattering in the time-domain model, as has been done in the frequency domain [25].

### C. Metamaterial on quartz substrate

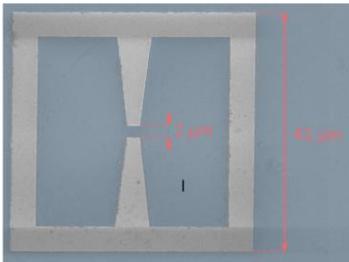

FIG. 13 SRR SEM image of the metasurface fabricated from e-beam lithography and lift off of gold evaporated onto a quartz substrate.

To go a step further, we used Fit@TDS on an artificially structured material called a metasurface. Metasurfaces are spreading their use in the THz range, for example as narrow-band terahertz modulators [55] or light-matter interaction enhancers [56]. Having precise insight of the properties of such components will accelerate the optimization of the fabrication process and allow the design of improved metasurfaces. Thus, we fabricated a metasurface made of a split-ring resonator (200 nm of Au and 20 nm of Ti) on top of a 200 µm-thick crystalline quartz substrate (z-cut) using electron-beam lithography and lift off. A scanning-electron microscope (SEM) image of the metasurface is shown in FIG. 13.

The filling factor of the quartz covered by metal is ~7%, corresponding to our previous hypothesis (see Sec. III.B for details). To perform the experiments, we tied the samples on top of a 1-mm-thick quartz substrate to prevent any interference between the near-field modes of the metasurface and the back interface of the thin substrate. The samples were then measured and fitted. As a first step the thickness and refractive index of the quartz without metasurface was retrieved to compare with the one found in the fit of the metasurface. Both results are shown in Fig. 14.

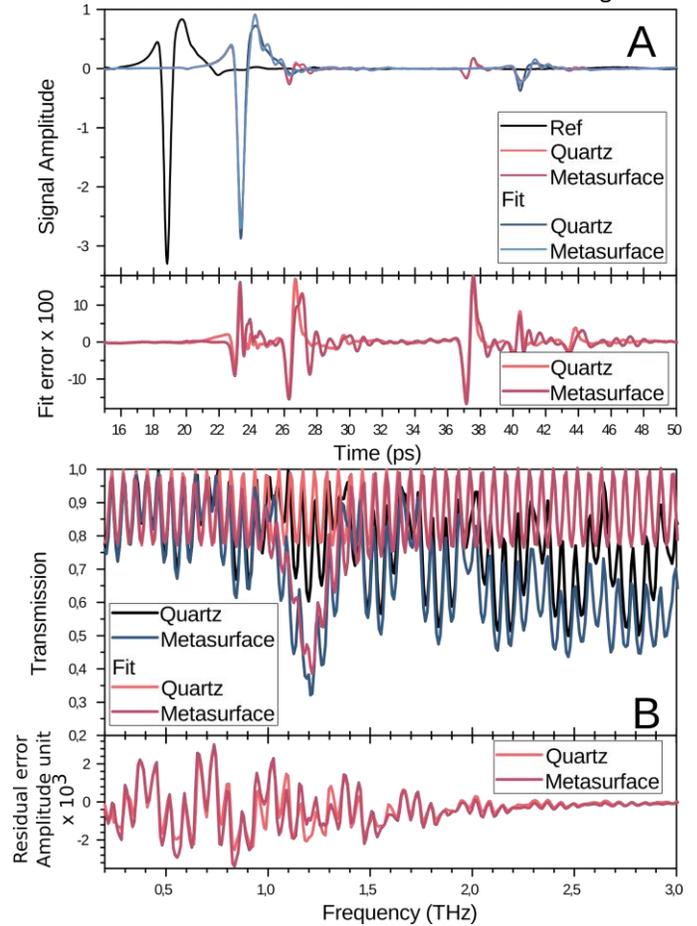

FIG. 14 (A) Top – time-trace of the result and the fit, bottom – the residual error; B) top – frequency-domain transmission, bottom – frequency-domain error (in amplitude units).

There, one can see that the fit succeeds with good precision (9.2% and 10.7% residual error, respectively). From the temporal residual error, we deduce that the small residual error arises from two different effects. First, two temporal pulses are not fitted by the model around 27 and 37 ps. Those two peaks correspond to unwanted reflections at the thick / thin quartz substrate, undoubtedly due to a thin air layer between them (the peak at 27 ps corresponds to propagation through ~ 211 µm of quartz, while the one at 37 ps to corresponds to 973 μm). This accounts for most of the residual error (~ 70%). Second, there remains a residual error that is temporally correlated with the main peak. No specific spectral feature is responsible for this residual error when analyzing the Fourier transform of this peak. We attribute this to the modification due to the previously reported effect. Indeed, since energy appears in other peaks, it must be removed from the main one.

Despite these effects, the software is able to retrieve the parameters of the metasurface as shown in Table 5.

We note that these parameters are coherent with the material's refractive index. Additionally, the total quality factor $Q_{tot}$ corresponds the one found by dividing the frequency by the width of the transmission deep. Both of these points confirm that the method produces reliable results. It also allows one to go a step further in making the difference between external losses (75%) and absorption losses (25%), which is close to the value usually found in simulations. Finally, since the metasurfaces are processed on a quartz/air interface, it is clearly non-symmetric. The coupling originating from the quartz is about a factor of three larger than that from air. This is again expected from momentum conservation, and corresponds to the depths of the observed peak as demonstrated in Ref. [57].

| Sample | Quartz | Metasurface |
|---|---|---|
| T (mm) | 1.2092 | 1.2101 |
| ε | 4.50 | 4.49 |
| $\omega_{meta}$ (THz) | | 1.220 |
| $Q_0$ | | 24.15 |
| $Q_{E1}$ | | 18.40 |
| $Q_{E2}$ | | 5.26 |
| $Q_{TOT}$ | | 6.11 |
| $\delta\theta$ (Rad.) | | 0.363 |

TABLE 5 : PARAMETERS FOR METASURFACE TDCMT FIT. THE QUALITY FACTOR IS DEFINED AS $Q = \omega\tau/2 \, \Delta T_{opt}$ GIVEN FOR SAKE OF USE; $Q_{TOT}$ IS CALCULATED AS IN REF. [34].

To summarize, Fit@TDS enabled us to retrieve all of the parameters of a resonant mode of a metasurface – showing that not only can material parameters be retrieved using our method but also photonic parameters. In addition, the residual error of the fit was clearly related to experimental perturbations, which can help a user to understand their experiment.

## VI. Conclusion

In this paper, we presented a new method, and an associated software, to fit THz-TDS data with material and metamaterial models based on time-domain fits. The goal of the software is to provide an improved, robust tool that gives more precise insight into an unknown sample, as well as to accelerate the analysis of a known sample, for instance, in the case of quality control. This software is freely available, and we provide a link to the source code in Ref. [31]. We first explained the methods, then tested it on fictitious samples and finally on real ones: a semiconductor, a molecular crystal and a metamaterial. Compared to other available software and methods, Fit@TDS has five main advantages. (*i*) A precise measurement of the thickness of the sample is not needed since it plays the same role as the other fit parameters. In fact, obtaining a precise sample thickness for materials such as carbohydrates or semiconductor wafers at sub-micrometer precision is challenging. Therefore, avoiding this step is a real improvement. (ii) We analyze the refractive index modelling problem as a whole, and thus we have only a small number of parameters for the fit compared to the usual two values per frequency, we are much less sensitive to the noise. This enables us to reach very high precision on the refractive index (< 10-3 for the silicon wafer, due to the set up limitations). (iii) Since the residual fit error is in amplitude units, one can clearly make interpretations of this error that lead to better understanding of experiments and possible oversights of the model implemented. In particular, this can reveal imperfections in the experiments, as we demonstrated on the quartz/metasurface sample. (*iv*) Since we are fitting in the time domain, the phase is not lost in the presence of strong absorption, and an additional step is not needed [23], as shown with our experiments on lactose. (*v*) Finally, it allows precise, reliable and consistent retrieval of material parameters using the Drude-Lorentz model, but also those of metamaterials with TDCMT—giving access to the internal and external losses of the metamaterial, as well as the coupling directivity.

Indeed, this method is not limited to THz-TDS systems and can be applied to any time-domain spectroscopic system if one has access to the electric or magnetic field. For instance, implementing this method with a dual-comb spectroscopy system (see Ref. [58] for a review of Asynchronous optical sampling ASOPS) would allow one to the follow of the fit parameters in real time at a 20 kHz rate. To further enlarge the scope of applications, it is possible to simply change the model in the open-source code of the software. It will then be possible to simulate other materials, for example, taking into account scattering [25], using Debye-derived models for liquid or impregnated samples [59], or even combine it further with mixture identification methods [60]. Alternatively, as a first step, one can simply implement the oblique incidence or the modeling of measurements using a reflection setup. Additionally, since the software enables one to simulate the photonic part, it will be possible to implement circuits made of THz resonators [61], or any other THz photonic component involving dispersive elements [53].

Since the residual error of the parameters reach the THz-TDS set up noise limitation, an improvement of the performance of the software would require an increase in the sensitivity of THz-TDS systems. For instance, several groups are working on producing emission antenna able to deliver more powerful and broadband pulses [62], as well as improved detection systems [63]. Furthermore, it would be very helpful to minimize the post-pulse emission of the antenna. This would allow the use of additional noise-reduction techniques [64], and thus improve the bandwidth and the precision of the fit.

To conclude, we hope that the community will make use of Fit@TDS and implement additional features corresponding to their needs. For example, one could imagine implementing a model to determine the concentrations of a known gas mixture, or of doping or impurity concentrations in a known material. Overall, because Fit@TDS functions on common personal computers and operating systems, we anticipate that it will become a valuable tool for the community and will help the spread of THz-TDS to new fields of research.